\documentclass[conference]{IEEEtran}
\usepackage{amssymb}
\usepackage{textcomp}
\DeclareUnicodeCharacter{2032}{\textprime}
\IEEEoverridecommandlockouts
\usepackage{cite}
\usepackage{amsmath,amssymb,amsfonts}
\usepackage{algorithmic}
\usepackage{graphicx}
\usepackage{xcolor}
\usepackage{booktabs}
\usepackage{pifont}
\begin{document}

\title{Exploring User Risk Factors and Target Groups for Phishing Victimization in
Pakistan\\
}

\author{\IEEEauthorblockN{\textsuperscript{} Javara A. Bukhsh}
\IEEEauthorblockA{\textit{SCS, EEMCS} \\
\textit{University of Twente}\\
Enschede, The Netherlands \\
j.a.bukhsh@utwente.nl}
\and
\IEEEauthorblockN{\textsuperscript{} Maya Daneva}
\IEEEauthorblockA{\textit{SCS, EEMCS} \\
\textit{University of Twente}\\
Enschede, The Netherlands \\
m.daneva@utwente.nl}
\and
\IEEEauthorblockN{\textsuperscript{} Marten van Sinderen}
\IEEEauthorblockA{\textit{SCS, EEMCS} \\
\textit{University of Twente}\\
Enschede, The Netherlands \\
m.j.vansinderen@utwente.nl}
}

\maketitle

\begin{abstract}
Phishing attacks pose a significant cybersecurity threat globally. This study investigates phishing susceptibility within the Pakistani population, examining the influence of demographic factors, technological aptitude and usage, previous phishing victimization, and email characteristics. Data was collected through convenient sampling; a total of $164$ people completed the questionnaire. Contrary to some assumptions, the results indicate that men, individuals over 25, employed persons and frequent online shoppers have relatively high phishing susceptibility. The characteristics of email significantly affected phishing victimization, with authority and urgency signaling increasing susceptibility, while risk cues sometimes improved vigilance. In particular, users were more susceptible to emails from communication services such as Gmail and LinkedIn compared to government or social media sources. These findings highlight the need for targeted security awareness interventions tailored to specific demographics and email types. A multi-faceted approach combining technology and education is crucial to combat phishing attacks.
\end{abstract}


\section{Introduction}
Phishing remains one of the most prevalent and damaging forms of cybercrime, with wide-ranging impact on both individuals and organizations \cite{vayansky2018phishing}. These attacks not only result in significant financial losses, but they also compromise personal data, break trust in digital platforms, and can lead to long-term psychological distress for victims \cite{abroshan2021covid,abroshan2021phishing, wang2023psychological, mahamood2023analysis, jaishankar2008identity}. From an organizational perspective, phishing is a significant threat regarding for data breaches, with the potential to damage reputations and disrupt business operations\cite{bose2014phishing}. The urgency of investigating phishing is further supported by statistics, such as the FBI IC3's 2023 Annual Report, released in 2024, shows a record 880,418 complaints totaling over \$12.5 billion in potential losses. This is a nearly 10\% increase in complaints and a 22\% increase in losses compared to 2022 \cite{FBI2024Annual}. Understanding phishing mechanics is essential for mitigating its impact and informing broader cybersecurity policies. 

This study examines how socioeconomic and cultural characteristics influence phishing susceptibility by investigating the Pakistani population as a case study. Pakistan presents an interesting context due to its distinct combination of rapid digital adoption (45.7\% internet penetration in 2024~\cite{Kemp2023DigitalPakistani}), significant youth demographic, and varying levels of digital literacy across different social groups. Understanding how these characteristics affect phishing susceptibility can help predict vulnerability patterns in populations with similar socioeconomic profiles.


Email phishing remains the most prevalent form of phishing, surpassing other types like vishing (voice phishing) and smishing (SMS phishing). Within email phishing, more sophisticated tactics such as spear phishing, CEO fraud, evil twin pharming, and angular phishing are frequently used. These techniques target specific individuals or organizations, often with highly personalized messages or fake networks that aim to deceive victims into revealing sensitive information, making email phishing a highly effective and dangerous threat in the cybersecurity landscape. 

A complex interplay of factors, including demographics, internet usage habits, and levels of IT awareness, influences phishing susceptibility.  Previous studies have examined these relationships across various populations~\cite{ribeiro2024factors, butavicius2017understanding}. However, conflicting findings suggest that socioeconomic and cultural factors significantly influence vulnerability patterns. For instance, studies show inconsistent results regarding gender-based susceptibility, with some finding higher female vulnerability~\cite{Sheng2010} while others indicate male users are more susceptible~\cite{Ong2014avast, hadlington2017human}, potentially due to differing cultural contexts and technology adoption patterns. This research examines these relationships within the Pakistani context, investigating how factors such as demographics, IT knowledge, internet usage patterns, prior phishing experiences, and persuasive message tactics interact with distinct cultural and socioeconomic characteristics. By analyzing these relationships in a rapidly digitalizing society with diverse digital literacy levels, this study contributes to developing a more nuanced, culturally-aware cybersecurity awareness programs for understanding and predicting phishing vulnerability across different populations that are similar to Pakistan. 


To assess phishing susceptibility, we presented participants with a combination of phishing and legitimate emails, evaluating their ability to accurately distinguish between them.  This study offer critical insights into factors that influence phishing susceptibility among the Pakistani population.

\section{Background and related work}
The term \textit{Phishing} describes how attackers use deceptive messages to lure victims into revealing personal information \cite{alabdan2020phishing}, with the ``pH" spelling influenced by ``phone phreaking" \cite{khonji2013phishing}. Phishing, a growing cybercrime with debated definitions \cite{lastdrager2014achieving, salloum2022systematic} emerged in 1996 from social engineering attacks. Lastdrager (2014)~\cite{lastdrager2014achieving} defines it as ``a scalable act of deception whereby impersonation is used to obtain information from a target" \cite{lastdrager2014achieving}. In addition, Phishing susceptibility is the likelihood of an individual becoming a victim of a phishing attempt \cite{chen2020examination}. This likelihood is higher if the phishing attempt is more sophisticated, e.g., by using a phishing email that better exploits human weaknesses. 


Beyond large-scale data breaches, phishing attacks silently compromise individuals' trust in technology and digital security, often causing lasting psychological distress.  The susceptibility to phishing is a multifaceted issue influenced by a combination of individual, situational, and technical factors \cite{chen2011assessing, wang2009visual}. The sophistication of the phishing attack itself, including the use of realistic branding, personalized content, and urgent language, can sometimes overwhelm even technically savvy individuals \cite{wash2020experts}.
 While global research on phishing susceptibility is extensive, studies focusing on the Pakistani context remain notably scarce. This study addresses this research gap by investigating how demographics, technological aptitude and usage patterns, repeated exposure to phishing attempts, and email characteristics influence phishing susceptibility within Pakistan's unique cultural and technological landscape. 
The following sections review the relevant literature on phishing susceptibility factors, leading to the formulation of this study's hypotheses.

\subsection{Demographics}
Although numerous studies explore the relationship between phishing susceptibility and individual demographics, a definitive consensus on the factors influencing susceptibility remains uncertain. For instance, a study including Australia, Canada, New Zealand, England, and the USA concluded no significant gender-based differences in phishing susceptibility  \cite{gopavaram2021cross}. Multiple studies support this finding~\cite{canfield2016quantifying, sarno2017phishers, moody2017phish, chen2020examination}. This contrasts with other research suggesting a relationship between gender and susceptibility to phishing attacks, highlighting a potential disagreement in the field \cite{sheng2010falls,abroshan2021phishing}. Some studies suggest that women are more vulnerable to phishing attacks \cite{jagatic2007social, goel2017got}, others indicate that men are more susceptible 
\cite{Ong2014avast, hadlington2017human}. The results are also influenced by the conceptualization of phishing susceptibility used in each study. Research supports that phishing susceptibility also relates to age, education, and profession. Recent studies have found that older people are more susceptible to phishing victimization due to a lack of digital knowledge \cite{lin2019susceptibility, gopavaram2021cross, o2021can}. In contrast, older studies stated that young people are more susceptible to phishing compared to the elderly population \cite{kumaraguru2007getting, sheng2010falls, darwish2012towards}. Some studies suggest that the level of education plays a vital role in phishing susceptibility. Individuals with higher levels of education may be better equipped to recognize and avoid phishing attempts due to increased awareness and critical thinking skills. In contrast, those with less formal education may be more vulnerable due to limited exposure to cyber threats and online safety practices \cite{moody2017phish,ghazi2022phishing}. Similarly, certain professions may be at higher risk due to their reliance on digital communication and sensitive data. For example, individuals in finance or healthcare may be specifically targeted due to the potential for high-value data breaches. Finally, profession can also be a factor, with certain professions potentially being targeted more than others\cite{gordon2019assessment, khasuntsev2024effectiveness, doing2024analysis}. However, specific research on the relationship between demographic and phishing susceptibility within the Pakistani population is currently lacking.

\subsection{Technological aptitude and usage}
Technology aptitude and usage significantly influence phishing susceptibility \cite{parsons2017human,albladi2018user,ayaburi2024technology}. Individuals with higher technology aptitude often demonstrate greater awareness of online security threats and are more adept at recognizing subtle indicators of phishing attempts\cite{ayaburi2024technology}. Their proficient usage of technology equips them with the skills to critically evaluate digital communications, identify inconsistencies, and avoid falling victim to deceptive tactics \cite{goel2017got, albladi2018user, ghazi2022phishing}. 

However, even tech-savvy individuals are not immune; risky online behaviors, such as visiting unverified websites or carelessly sharing personal information, can significantly increase susceptibility \cite{gan2024fishing}. Internet usage for essential activities like online banking, social media, or work/study presents a nuanced risk landscape. Regular engagement can increase familiarity with online interfaces, but simultaneously exposes users to a greater volume of potential phishing attempts \cite{downs2006decision, sarno2022so}. Navigating this complex landscape requires a constant vigilance and a proactive approach to online security.
\subsection{Repeat phishing experience}
Repeated phishing experiences can significantly impact an individual's future susceptibility and online behavior. Research presents mixed findings on the causes of past phishing victimization. Some studies suggest that prior victims may not readily learn from their experiences, easily recover from the impact of victimization, and exhibit a lack of vigilance when subsequently exposed to phishing attacks \cite{correia2020patterns,caputo2013going}. Other research indicates a correlation between certain online behaviors, such as frequent online shopping and previous victimization, and an increased susceptibility to phishing scams in future\cite{de2018you,li2020experimental}. However, some studies contradict this, suggesting that previous victimization  make individuals more careful,  lead to safer online habits or practice safety more often regarding phishing attacks\cite{hassandoust2020role, chen2020examination}. This discrepancy highlights the complexity of phishing susceptibility, which can be influenced by a range of individual, situational, and technical factors.
\subsection{Decision factors}
When individuals encounter phishing emails, their decision to respond is often influenced by a complex interplay of factors. Research suggests that elements such as perceived authority, perceived risk, and the conveyed sense of urgency play significant roles in shaping this decision-making process\cite{morrow2024scamming,bayl2022response}. The presence of authority cues, such as official-looking logos or claims of representing legitimate organizations, can increase the likelihood of compliance \cite{ferreira2019persuasion, tiwari2020exploring}. Furthermore, the use of urgency and authority tactics, designed to create a sense of immediate action required, can override rational assessment and lead to impulsive responses \cite{PhishMe2017Human, williams2018exploring}. Conversely, the perception of risk, including concerns about potential financial loss, identity theft and loss of reward may deter individuals from engaging with the email \cite{goel2017got, hawamdah2024hooks}. Understanding the relative weight and interaction of these decision factors – authority, urgency and risk – are crucial for developing effective strategies to mitigate the impact of phishing attacks.
\subsection{Source of emails}
The perceived credibility of the email source significantly impacts an individual's susceptibility to phishing attacks. Emails purportedly originating from trusted entities, such as government institutions or well-known communication platforms, may engender a higher degree of trust, leading users to overlook potential red flags. Recent advisories from Pakistan's National Cyber Emergency Response Team (PKCERT)\cite{Sarfraz2022FBR, Ali2025Pakistani} highlight the ongoing threat of sophisticated phishing attacks targeting Pakistani citizens and critical infrastructure. These attacks often leverage the authority of government institutions, with emails falsely claiming to be from the "Office of Commissioner Police Department" or using government job advertisements to acquire confidential information.
 Social media platforms, with their inherent emphasis on social connection, present a unique avenue for phishing, as users may be more likely to trust messages appearing to come from their network or from the platform itself. Therefore, understanding the interplay between source credibility, user expectations, and phishing techniques is crucial for developing effective detection and prevention strategies. 


\subsection{Phishing susceptibility in Pakistan}
 Given this uncertainty, Pakistan faces significant challenges regarding phishing susceptibility due to various  factors. First, the youth, who constitute the largest portion of internet users, are particularly vulnerable. While precise, up-to-date figures are difficult to obtain, a 2023 report indicated that 36.7\% of the total population were internet users, and 82.1\% of those users were on social media \cite{Kemp2023DigitalPakistani}. This suggests a large number of young Pakistanis are online, potentially without adequate cybersecurity awareness. This lack of awareness is being addressed by National CERT, which has launched a cybersecurity awareness campaign but only targeting children \cite{PKCERT2025Junior}. Second, the level of education plays a crucial role. A large segment of the Pakistani population has limited education \cite{Macrotrends2025PKliteracy}, which is correlated with limited internet usage \cite{Kemp2023DigitalPakistani}. This lack of digital literacy can make individuals more susceptible to phishing attacks, as they may not be able to recognize red flags or understand the risks involved. Thirdly, gender has been shown to influence susceptibility to phishing. One study found that women in Pakistan do experience cybercrime; however, a lack of cybersecurity awareness, combined with geographic and societal barriers, often makes it difficult for them to report incidents or access support services \cite{tarar2021cybercrimes}.  Despite such findings, there is a significant gap in research that specifically examines how demographic factors such as age, gender, education, and profession influence the susceptibility to phishing within the Pakistani population. To date, no comprehensive studies have been conducted on this subject, making this research among the first to explore these dimensions in the local context. 

 \subsection{Hypothesis}
 Based on the overview of related work presented above, we formulate the following hypotheses to examine phishing susceptibility patterns within the Pakistani population:
\begin{itemize}
    \item[\textbf{H1:}]Female are more susceptible to phishing attacks than male. \par
    
    \item[\textbf{H2:}] Younger age groups (17-24) are more susceptible to phishing attacks than older age groups. \par
    
    \item[\textbf{H3:}] Education level 
    is negitively associated with phishing susceptibility.\par
    
    \item[\textbf{H4:}] Students demonstrate higher phishing susceptibility compared to employed individuals. \par
    
    \item[\textbf{H5:}] Years of internet experience is positively associated with phishing susceptibility. \par
    
    \item[\textbf{H6:}] Time spent per week on online shopping and banking is positively associated with phishing susceptibility. \par
    
    \item[\textbf{H7:}] Higher levels of self-reported IT knowledge are positively associated with phishing susceptibility. \par
    
\item[\textbf{H8:}] Individuals with better email security awareness (e.g., being cautious of unknown senders) demonstrate lower phishing susceptibility. \par
    
    \item[\textbf{H9:}] Previous phishing victims are more likely to fall for phishing again.\par
    
    \item[\textbf{H10:}] Emails containing authority cues increase phishing susceptibility. \par
    
    \item[\textbf{H11:}] Emails containing urgency cues increase phishing susceptibility. \par

   \item[\textbf{H12:}] Emails containing risk factor cues result in higher phishing susceptibility. \par 
    
    \item[\textbf{H13:}] Phishing emails purporting to be from social media sources will result in lower user susceptibility rates.\par
  \item[\textbf{H14:}] Phishing emails claiming to originate from communication services will generate a lower user susceptibility. \par
 \item[\textbf{H15:}] Phishing emails that appear to come from government organizations will elicit higher susceptibility. \par

\section{Research methodology}
A model of phishing susceptibility (Fig.1) was developed to guide the study. The model posits that demographic factors, technological aptitude, repeat phishing victimisation, and email-related factors influence phishing susceptibility. The questionnaire was designed to collect data on these factors. This section describes our questionnaire design, data collection process, and measures used to assess the dependent and independent variables.
\subsection{Questionnaire}
The questionnaire\footnote{The questionnaire utilized in this study is available upon request.} comprises five categories. The first category, \textit{demographics}, includes age, gender, education, and profession to identify which demographic are more vulnerable to phishing victimization. The second category, \textit{technological aptitude and usage}, included internet usage, IT knowledge (IT knowledge was operationalized through self-reported measures, where participants rated their familiarity with technology on a scale from ``not aware" to ``very aware") and human aspects (by using Human Aspects of Informataion System - Questionniare (HAIS-Q) framework \cite{parsons2017human}. This aims to assess how time spent on the internet, knowledge of IT, and individual understanding of email usage influence phishing vulnerability.   The third category, \textit {repeat phishing experience}, evaluates how much an individual learned from previous phishing encounter(s) and associated losses. The fourth category, \textit {decision factors} examines how urgency, authority, and risk factors influence response behaviors. These persuasion factors influence rational decision-making regarding how to react on the email\cite{jansen2018persuading,cialdini2001science}. The fifth category, \textit {source of email}, presents both phishing and legitimate emails from various sources including social media,  communication and different government institutes in Pakistan. This category assesses how email source affects response behavior.  We used feedback from two pilot participants to refine the questionnaire before a broader distribution.
\subsection{Data collection}
We used Microsoft Forms to survey participants in Pakistan from October 2024 to February 2025, distributing the survey through social networks (Facebook and Instagram). On average, the survey took 15 minutes to complete.
\subsection{Measures}
The following section explains how the dependent variable, independent variables, and email characteristic are measured and encoded for analysis.
\subsubsection{Dependent variable}

Phishing susceptibility served as the dependent variable in this study. Participants were presented with a total of 43 emails ($28$ phishing and $15$ legitimate) and asked to categorize each email into one of two types: phishing or legitimate email. We computed an overall susceptibility score based on a participant's accuracy in identifying phishing emails, with particular focus on \textit{false negatives} (where phishing emails were incorrectly labelled as legitimate). The susceptibility threshold was set at the median of false negative scores, with participants scoring above the median classified as susceptible ($1$) and those below as not susceptible ($0$).

\subsubsection{Independent variables}
The independent variables in this study encompassed demographics, technology aptitude, and previous phishing experience. Demographic variables included gender (coded as 0 = male, 1 = female), age (0 = 25 and above, 1 = 17-24), education level (ranging from 1 to 6), and professional status (0 = employed, 1 = student). Technology aptitude was assessed through multiple measures: internet use experience (scaled from 1 = less than 1 year to 8 = more than 20 years), weekly time spent on online activities such as shopping, banking, or social media (ranging from 0 = none to 6 = more than 8 hours), self-assessed IT knowledge (scaled from 1 = not aware to 5 = very aware), and security awareness measured using HAIS-Q (Likert scale: 1 = strongly agree to 5 = strongly disagree). The previous phishing experience was captured through two key variables: whether the participants had fallen victim to phishing attacks (coded 1 = yes, 0 = no/don't know) and the consequences they experienced, including monetary loss, identity theft, ransomware attacks, and reputation damage (each coded 1 = experienced, 0 = not experienced). These variables were selected to comprehensively capture participants' technological background, previous exposure to phishing attacks, and potential susceptibility factors.

\subsubsection{Email Characteristics}
The 43 emails presented to participants were systematically designed to incorporate various characteristics and sources, for analysis. Each email was categorized based on its source (social media, communication services, or government organizations) and had one of three persuasion cues: authority, urgency and risk level. The participant was asked to classify the email as legitimate or phishing. 
For each participant-email combination, we recorded the participant's response (correct or incorrect identification), with particular attention to false negatives (where phishing emails were incorrectly identified as legitimate). We then calculated participant-level metrics, including an overall accuracy score and a false negative count.




\begin{figure}
    \centering
    \includegraphics[width=\linewidth]{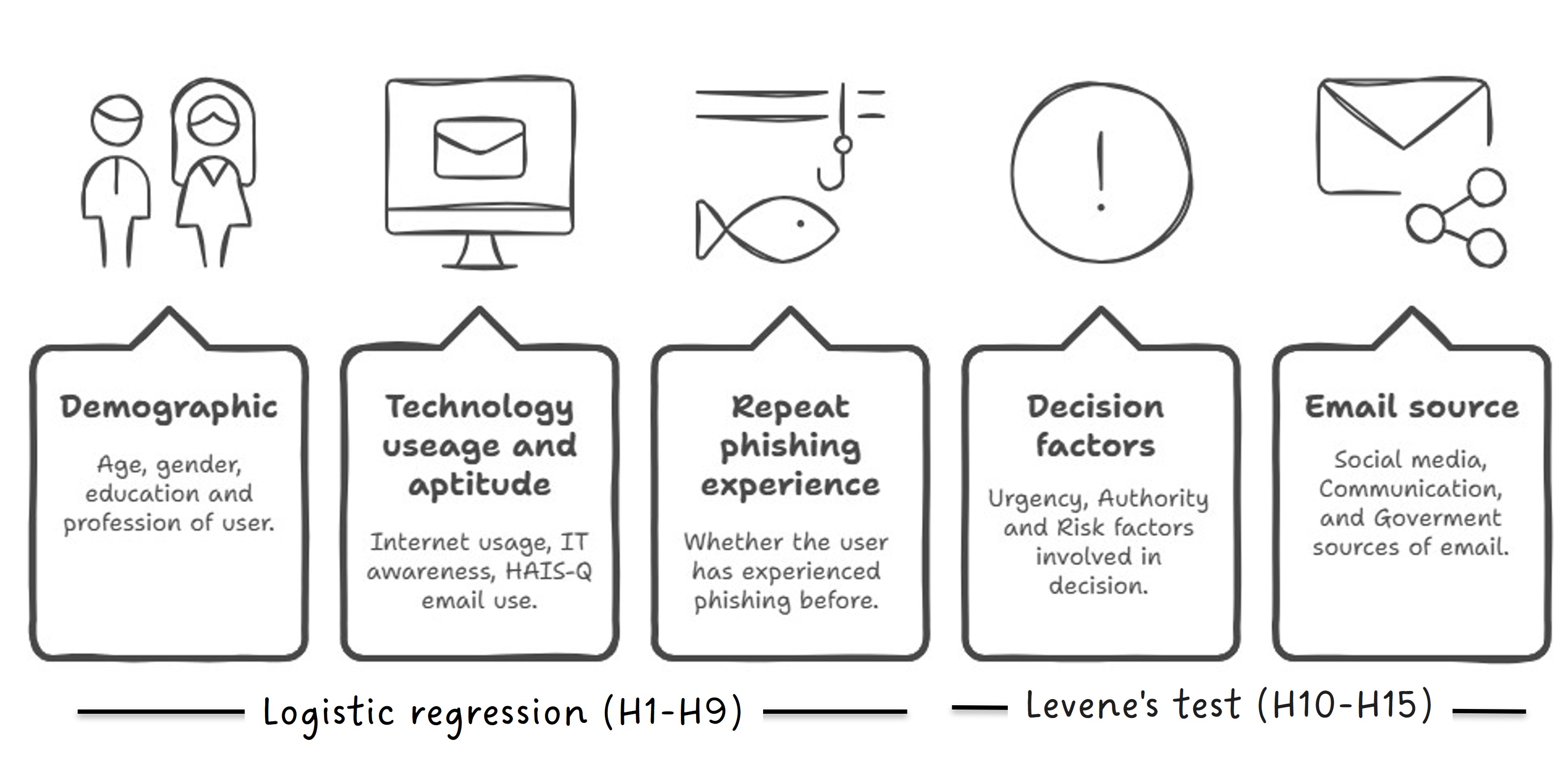}
    \caption{Model of Phishing Susceptibility}
    \label{fig:enter-label}
\end{figure}

\end{itemize}


\section{Results}
\subsection{Individual Characteristics and Phishing Susceptibility}
We examined how individual characteristics influence phishing susceptibility using logistic regression analysis (N = 164, H1-H9). This analysis approach aligns with methods employed in similar phishing research studies with comparable sample size of  $145$ and $449$ \cite{ribeiro2024factors, abroshan2021phishing}. The analysis investigated demographic factors, technological aptitude, and previous phishing experience as predictors of participants' tendency to incorrectly classify phishing emails as legitimate. Table~\ref{tab:logit_H1_H9} presents the logistic regression results for the hypotheses tested.

\begin{table}[!htbp]
\centering
\caption{Logistic Regression Results Predicting Phishing Susceptibility (Combined Score)}
\label{tab:logit_H1_H9}
\begin{tabular}{llllc}
\toprule
Hypothesis &  $\beta$ (Signif.)  &   Odds Ratio   & 95\% CI   & Supported?       \\
\midrule
H1      & $-0.9389^{**}$  & 0.391   & [0.206, 0.741]    & \ding{55}  \\
H2     & $-0.9343^{**}$  & 0.3928  & [0.205, 0.752]  & \ding{55}   \\
H3      & $0.1823$        & 1.199   & [0.814, 1.767] & \ding{55}    \\
H4          & $-0.8119^{*}$   & 0.444   & [0.230, 0.856]  & \ding{55}   \\
H5      & $0.0509$        & 1.0522  & [0.868, 1.275] & \ding{55}    \\
H6          & $0.3736^{***}$  & 1.4528  & [1.165, 1.811] &\ding{51}    \\
H7      & $-0.1936$       & 0.8239  & [0.630, 1.078]  & \ding{55}   \\
H8          & $-0.589^{*}$    & 0.555   & [0.309, 0.996]  &\ding{51}   \\
H9          & $0.827^{*}$     & 2.290   & [1.080, 4.830]   & \ding{51}  \\
\bottomrule
\multicolumn{4}{@{}l@{}}{\footnotesize Note: *p $<$0.05, **p $<$ 0.01, ***p $<$ 0.001}
\end{tabular}
\end{table}

Our analysis of demographic factors revealed unexpected patterns in phishing susceptibility.  Regarding gender differences (H1), our analysis showed a significant effect of gender ($\beta$ = $-0.9389$, $p$ = $0.004$), but in the opposite direction than hypothesized. The negative coefficient indicates that females were less susceptible to phishing attacks than males. The odds ratio of 0.391 (95\% CI [0.206, 0.741]) indicates that females had 60.9\% lower odds of falling for phishing attempts compared to males. Therefore, H1 was not supported, with results showing males were significantly more susceptible to phishing attacks. Similar age based analysis (H2)  showed a significant effect of age ($\beta$  = -$0.9343$, $p$ = $0.005$), but in the opposite direction than hypothesized. The negative coefficient indicates that younger individuals (17-24) were less susceptible to phishing attacks than older individuals (25 and above). The odds ratio of 0.3928 (95\% CI [0.205, 0.752]) indicates that younger individuals had 60.72\% lower odds of falling for phishing attempts compared to older age groups. Therefore, H2 was not supported, with results showing older age groups were significantly more susceptible to phishing attacks. This finding challenges common assumptions about youth susceptibility to cyber threats and suggests a potential generational advantage in phishing detection.

Education level and professional status revealed varying influences on phishing susceptibility. Education level (H3) hypothesized a negative association with phishing susceptibility, where higher education levels would correspond to lower susceptibility. However, our analysis revealed no statistically significant relationship ($\beta$ = $0.1823$, $p$ = $0.356$). The odds ratio and wide confidence interval indicate no clear effect of education level on phishing susceptibility. Therefore, H3 was not supported. Similarly, our analysis for professional status (H4) showed a significant effect ($\beta$ = $-0.8119$, $p$ = $0.015$), but in the opposite direction than hypothesized. The negative coefficient indicates that students were less susceptible to phishing attacks than employed individuals. The odds ratio of 0.444 (95\% CI [0.230, 0.856]) indicates that students had $55.6$\% lower odds of falling for phishing attempts compared to employed individuals. Therefore, H4 was not supported, with results showing employed individuals were significantly more susceptible to phishing attacks.

Investigation of online experience metrics yielded contrasting results. For internet experience (H5), we hypothesized a positive association with phishing susceptibility, where more years of internet experience would correspond to higher susceptibility. Our analysis revealed no statistically significant relationship ($\beta$ = $0.0509$, $p$ = $0.604$). The odds ratio of 1.0522 (95\% CI [0.868, 1.275])  indicate no clear effect of internet experience on phishing susceptibility. Therefore, H5 was not supported, suggesting that years of internet experience alone does not significantly influence phishing susceptibility. However, regarding time spent on online shopping and banking (H6), our analysis revealed a significant positive association with phishing susceptibility ($\beta$ = $0.3736$, $p$ = $0.001$). Our result indicates that more time spent online is associated with higher phishing susceptibility. Therefore, H6 was supported, demonstrating that individuals who spend more time on online shopping and banking are significantly more susceptible to phishing attacks. 

Regarding self-reported IT knowledge (H7), we hypothesized a positive association with phishing susceptibility, where higher levels of self-assessed IT knowledge would correspond to higher susceptibility. Our analysis revealed no statistically significant relationship ($\beta$ = $-0.1936$, $p$ = $0.158$). The odds ratio of $0.8239$ (95\% CI [0.630, 1.078]) indicate no clear effect of self-reported IT knowledge on phishing susceptibility. While the negative coefficient suggests a trend toward lower susceptibility with higher IT knowledge, this trend was not statistically significant. Therefore, H7 was not supported, indicating that self-assessed IT knowledge does not reliably predict phishing susceptibility.

Both psychological and experiential factors played a
notable role. Regarding email security awareness (H8), we hypothesized that individuals with better security awareness would less likely to fall for phishing. Our analysis revealed a significant relationship between email security awareness and phishing susceptibility ($\beta$ = $-0.589$, $p$ = $0.048$).  The odds ratio of $0.555$ (95\% CI [0.309, 0.996]) indicates that for each unit increase in security awareness score, the odds of falling for phishing attempts decreased by $44.5$\%.  Therefore, H8 was supported, demonstrating that individuals with better email security awareness are significantly less susceptible to phishing attacks. Similarly, for previous phishing victimization (H9), our analysis revealed a significant positive association with phishing susceptibility ($\beta$ = $0.827$, $p$ = $0.030$). The positive coefficient indicates that previous victims were more susceptible to future phishing attempts. The odds ratio of 2.290 (95\% CI [1.080, 4.830]) indicates that previous phishing victims had  higher odds of falling for phishing attempts compared to those who had not been victimized before. Therefore, H9 was supported, demonstrating that previous phishing victimization is associated with increased susceptibility to future attacks.

\subsection{Email Characteristics and Phishing Susceptibility}
The impact of email characteristics on phishing susceptibility was analyzed using Levene's test, examining both persuasion cues (authority, urgency, and risk factors) and purported sources (social media, communication services, and government organizations). This analytical approach aligns with the methodology reported in ~\cite{tiwari2020exploring}, where the author used Levene's test to examine the effect of email characteristics on phishing susceptibility. Unlike participant characteristics, these features varied across emails, resulting in different numbers of observations for each condition, making this method particularly suitable for analyzing within-subject variations in response to different email characteristics. Table~\ref{tab:logit_H10_H15} provides results of descriptive statistics and Levene's test for hypotheses.

\begin{table*}
\caption{Levene's Test Results for Email Characteristics and Phishing Susceptibility}
\label{tab:logit_H10_H15}
\centering
\small
\begin{tabular}{@{}llcccccc@{}}
\toprule
  Cue/Source & Present & Absent & N & Mean & Levene's & Effect & Hypothesis/Supported? \\
 &  & (M $\pm$ SD) & (M $\pm$ SD) & (Present/Absent) & Difference & Test & Size \\
\midrule
H10 /\ding{51} & Authority & 0.693 $\pm$ 0.461 & 0.629 $\pm$ 0.483 & 2624/1968 & 0.064 & 20.882*** & -0.064   \\
H11/ \ding{51} & Urgency & 0.695 $\pm$ 0.460 & 0.529 $\pm$ 0.499 & 3772/820 & 0.166 & 85.036*** & -0.166 \\
H12  /\ding{55} & Risk factors & 0.636 $\pm$ 0.481 & 0.678 $\pm$ 0.467 & 1312/3280 & -0.042 & 7.463** & 0.042  \\
H13/ \ding{51} & Social media & 0.615 $\pm$ 0.487 & 0.690 $\pm$ 0.463  &  1476/3116 & -0.074 & 25.099*** & 0.074   \\
H14  /\ding{55} & Communication & 0.741 $\pm$ 0.439 & 0.630 $\pm$ 0.483 & 1476/3116  & 0.110 & 55.311*** & -0.110   \\
H15  /\ding{55} & Government & 0.644 $\pm$ 0.479 & 0.678 $\pm$ 0.467 & 1640/2952  &  -0.034 & 5.462** & 0.034 \\

\bottomrule
\multicolumn{8}{@{}l@{}}{\footnotesize Note: **p $<$ 0.01, ***p $<$ 0.001}
\end{tabular}
\end{table*}

Phishing preys on individuals by leveraging authority, urgency, and personal risk factors to bypass critical thinking. To investigate it further H10 stated that emails containing authority cues increase phishing susceptibility. Results support this hypothesis and indicated that emails containing authority elements showed significantly higher susceptibility rates (M = 0.693, SD = 0.461) compared to those without (M = 0.629, SD = 0.483), with a significant Levene's test result (F = 20.882, p $<$  0.001) and a small negative effect size (-0.064). Related to urgency factors, H11 stated that emails containing urgency cues increase phishing susceptibility. Urgency cues demonstrated the strongest impact among all characteristics examined, with emails containing urgency elements eliciting substantially higher susceptibility (M = 0.695, SD = 0.460) compared to those without (M = 0.529, SD = 0.499; F = 85.036, p $<$ 0.001, effect size = -0.166). Focusing on risk factors, H12 proposed that emails containing risk factor cues result in higher phishing susceptibility. Risk factor-based cues did not increase susceptibility; emails containing risk factor elements actually showed slightly lower susceptibility rates (M = 0.636, SD = 0.481) compared to those without (M = 0.678, SD = 0.467; F = 7.463, p $<$ 0.01).

Analysis of email sources revealed varying patterns of susceptibility. H13 poposed that phishing emails purporting to be from social media sources will result in lower user susceptibility rates. Social media-sourced emails resulted in lower susceptibility rates (M = 0.615, SD = 0.487) compared to other sources (M = 0.690, SD = 0.463; F = 25.099, p $<$0.001), with an effect size of 0.074. This suggests that users may be more cautious when evaluating emails claiming to be from social media platforms. \par According to H14, phishing emails claiming to originate from communication services will generate a lower user susceptibility. Supporting H14, emails purporting to be from communication services generated significantly higher susceptibility rates (M = 0.741, SD = 0.439) compared to other sources (M = 0.630, SD = 0.483; F = 55.311, p $<$ 0.001), with a notable effect size of -0.110. \par
Lastly, H15 proposed that phishing emails that appear to come from government organisations will elicit higher susceptibility. Contrary to H15, government-sourced emails did not elicit higher susceptibility rates (M = 0.644, SD = 0.479) compared to other sources (M = 0.678, SD = 0.467; F = 5.462, p $<$ 0.01).

These findings suggest that urgency and authority cues are particularly effective in increasing phishing susceptibility, while risk cues may actually enhance users' vigilance. Among source types, communication service-themed emails pose the greatest risk, challenging assumptions about the relative dangers of government or social media-themed phishing attempts.

\section{Discussion}
Our findings reveal a complex interplay of various factors, highlighting key phishing susceptibilities within the Pakistani context.
Our results contradict H1, which posited a higher female susceptibility to phishing. Contrary to some previous research, our data indicate that males in the Pakistani population are more vulnerable. Our finding is consistent with studies such as \cite{Ong2014avast, hadlington2017human}. Hadlington et al. 
 (2027) \cite{hadlington2017human} attributes that phishing susceptibility to men, showing more trust and comfort when using personal devices.
 Our study revealed that individuals over 25 year age in Pakistan are more susceptible to phishing attacks, a finding supported by previous research, \cite{gopavaram2021cross, lin2019susceptibility, o2021can}, which attributes this increased susceptibility to a lack of education. In our study, we found no statistically significant association between education level and phishing susceptibility (H3). This result may stem from our specific sample characteristics or limitations in our measures of education and phishing susceptibility. A larger sample or more measures may reveal a subtle relationship. 
 Our findings confirmed the H4, indicating that students are less susceptible to phishing attacks compared to employed individuals. This difference may stem from employee's constant exposure to internet, cognitive overload, making them easier targets for phishing schemes. Risky online behavior may also contribute to this increased susceptibility \cite{downs2007behavioral,cuchta2019human, mohebzada2012phishing}.

Our findings indicate no significant relationship between phishing susceptibility and either years of internet experience (H5) or self-reported IT knowledge (H7) in the Pakistani population.
Regarding H6, we found a positive association between time spent per week on online shopping, banking and phishing susceptibility. Previous research is in line with our findings, noting that each transaction on a banking app or online shopping creates an opportunity for phishers to exploit vulnerabilities \cite{ghazi2022phishing, ribeiro2024factors}. Furthermore, we found that individuals with better email security awareness (e.g., being cautious of unknown senders) demonstrate a lower susceptibility to phishing (H8). This finding is supported by research showing that higher levels of email-specific Information Security Awareness (ISA) – encompassing knowledge, attitude, and behavior – are associated with better detection of both phishing and spear-phishing emails \cite{butavicius2017understanding}. Consistent with our hypothesis (H9), our findings indicate that previous phishing victimization does not significantly reduce susceptibility of Pakistani individuals to future attacks. Studies found that more than half of the previously phished individuals failed to identify the deceptive emails used.
This persistent susceptibility may stem from overconfidence in their ability to protect themselves from phishing attacks\cite{hong2013keeping, correia2020patterns, caputo2013going}. \par

Regarding H10, Our data supports that Pakistani individuals are susceptible to phishing emails containing authority cues. Several studies suggest that the presence of official-looking logos or claims of representing legitimate organizations can increase the likelihood of compliance \cite{ferreira2019persuasion, tiwari2020exploring}. This heightened susceptibility may stem from a desire to avoid potential negative consequences associated with non-compliance.
A parallel pattern emerged with respect to urgency factors in emails, with individuals demonstrating an increased susceptibility to these cues (H11). Studies indicate that offenders frequently employ urgency tactics to shape users' decision-making processes and capitalize on the situation, thereby increasing their susceptibility to phishing attacks \cite{morrow2024scamming, bayl2022response}. 
Contrary to H12, which predicted that emails containing risk factor cues would result in lower phishing susceptibility, the indiviuals showed no increased susceptibility to emails containing financial or other risk factors. These results contradict previous studies which found that people typically respond impulsively to emails containing financial themes, following attackers' requests due to fear\cite{hawamdah2024hooks, PhishMe2017Human, williams2018exploring}. \par

Regarding email sources, our data indicates that the individuals exhibit lower susceptibility to phishing emails originating from social media platforms (H13). While existing studies \cite{albladi2018user, mouncey2025phishing} extensively discuss social media phishing, limited research investigates susceptibility to emails mimicking social media notifications. This lower susceptibility may be because users typically interact with social media platforms through apps or websites rather than through email communications.


Conversely, our findings reveal higher susceptibility to emails purporting to be from communication services such as Gmail and LinkedIn (H14).  
This increased susceptibility may stem from users' concerns about the potential consequences of ignoring these emails, such as losing account access or missing important professional communications.
Contrary to H15, our data revealed lower phishing susceptibility with emails appearing to be from government institutions. This unexpected resistance may be attributed to widespread public awareness campaigns by government institutions and frequent media coverage in newspapers detailing phishing attacks and detection methods \cite{Sarfraz2022FBR, Ali2025Pakistani}. \par

Our research challenges simplistic assumptions about phishing vulnerability, revealing unexpected patterns of susceptibility and resilience across different email sources in the Pakistani population. These findings emphasize the importance of continuous monitoring and adaptation of security measures to address evolving threats and user behaviors.

\section{Conclusion and Limitations}
Through this comprehensive study, we found significant implications on phishing susceptibility within the Pakistani population that yielded several key insights that challenge common assumptions and advance our understanding of this critical cybersecurity issue.  \par
Our findings challenge conventional wisdom by demonstrating that males, older adults, employed individuals and those frequently engaged in online financial activities exhibit heightened vulnerability. The influence of email characteristics was also pronounced, with authority and urgency cues significantly increasing susceptibility, while risk cues unexpectedly promoted vigilance. Interestingly, phishing emails purporting to be from Gmail and LinkedIn elicited higher susceptibility among Pakistani users than those mimicking government or social media platforms. \par 
These findings have important implications for designing targeted security awareness interventions in Pakistan. Generic cybersecurity training may be insufficient; instead, tailored programs that address the specific vulnerabilities identified in this study are needed. For example, campaigns should focus on raising awareness about the deceptive tactics used in phishing emails and emphasize the importance of critically evaluating requests, even when they appear to come from trusted sources. Moreover, efforts to promote email security awareness should target specific demographic groups, such as older adults and frequent online shoppers, who are at higher risk.\par

Ultimately, this research underscores the importance of a multi-faceted approach to combating phishing, combining technological solutions with targeted education and awareness initiatives. By understanding the specific factors that influence phishing susceptibility within the Pakistani context, we can develop more effective strategies to protect individuals and organizations from these pervasive cyber threats. \par

This study makes a significant contribution to understanding phishing susceptibility in Pakistan by examining a wide range of demographic factors, internet usage habits, and psychological influences across various email sources. Unlike previous research \cite{ribeiro2024factors, downs2007behavioral} that often focused on a limited number of phishing emails, this study employed a comprehensive set of 43 phishing and legitimate emails. Furthermore, this is the first study on the Pakistani population that checks demographic susceptibility and covers decision factors and the impact of different sources of emails on Pakistani people. However, several limitations warrant consideration.\par
First, the sample size of $164$ participants is relatively small, which may limit the generalization of the findings to the broader Pakistani population. In addition, the reliance on convenience sampling may introduce selection bias, affecting the representativeness of the sample. In addition, the reliance on convenience sampling may introduce selection bias, affecting the representativeness of the sample. Future research should aim to replicate this study with a larger, more diverse sample to enhance the robustness and representativeness of the results. Second, the methodology relied on self-reported data and simulated phishing scenarios. A more valid approach would involve sending real (but ethically controlled) phishing emails to participants and tracking their actual responses. This would require obtaining email addresses from a large pool of individuals and monitoring their interactions, which presents significant logistical and ethical challenges.\par
To further advance this line of inquiry, several avenues for future research are recommended. A more generalized study at the provincial or institutional level would provide valuable insights into regional variations in phishing susceptibility. Additionally, replicating this study in other countries would help to identify cross-cultural differences and develop more universally applicable security awareness strategies, as individual country-level data remains limited. Finally, future research could explore the effectiveness of different intervention strategies, such as targeted training programs or technological solutions, in mitigating the specific vulnerabilities identified in this study, particularly those related to communication service-based phishing attacks.

\vspace{12pt}
\bibliographystyle{ieeetr}
\bibliography{mybibliography}
\end{document}